\documentclass[aps,pra,showpacs,twocolumn,superscriptaddress]{revtex4}
\usepackage{graphicx}
\usepackage{amsmath,amssymb,amsthm,dsfont,bm}

\begin{document}
\preprint{}
\title{Nonclassical polarization dynamics in classical-like states}
\author{Alfredo Luis}
\email{alluis@fis.ucm.es}
\homepage{http://www.ucm.es/info/gioq}
\affiliation{Departamento de \'{O}ptica, Facultad de Ciencias
F\'{\i}sicas, Universidad Complutense, 28040 Madrid, Spain}
\author{\'{A}ngel S. Sanz}
\affiliation{Instituto de F\'{\i}sica Fundamental (IFF-CSIC),
Serrano 123, 28006 Madrid, Spain}

\date{\today}

\begin{abstract}
Quantum polarization is investigated by means of a trajectory picture
based on the Bohmian formulation of quantum mechanics.
Relevant examples of classical-like two-mode field states are thus
examined, namely Glauber and SU(2) coherent states.
Although these states are often regarded as classical, the analysis
here shows that the corresponding electric-field polarization
trajectories display topologies very different from those expected
from classical electrodynamics.
Rather than incompatibility with the usual classical model, this result
demonstrates the dynamical richness of quantum motions, determined by
local variations of the system quantum phase in the corresponding
(polarization) configuration space, absent in classical-like models.
These variations can be related to the evolution in time of the phase,
but also to its dependence on configurational coordinates, which is
the crucial factor to generate motion in the case of stationary states
like those here considered.
In this regard, for completeness these results are compared with those
obtained from nonclassical $N00N$ states.

\end{abstract}

\maketitle


\section{Introduction}
\label{sec1}

According to its standard definition \cite{bornwolf-bk}, light
polarization refers to the ellipse described in time by the real
component of the electric-field vector of a harmonic wave.
Hence partial polarization can then be understood as the rapid and
random succession of more or less different polarization states.
In the quantum realm, we find that the electric field can never display
a well-defined ellipse, just in the same way that particles cannot
follow definite trajectories \cite{S1,S2,S3,S4,S5}.
This is because the (field) quadratures satisfy the same commutation
relations of position and linear momentum, and brings in several
remarkable consequences:
(i) there is no room for the classic, textbook definition of
polarization,
(ii) the simple and elegant picture of partial polarization as a
random succession of definite ellipses gets lost,
and (iii) any quantum light state is partially polarized because of
unavoidable (quantum) fluctuations.
The purpose of this work is to investigate whether these inconvenient
quantum consequences can be overcome resorting to the Bohmian picture
of quantum mechanics.

Polarization is a preferential laboratory for the analysis and
application of fundamental aspects of the quantum theory.
In this regard, one can get profit from tools coming from the latter
to analyze optical behaviors.
This is the case, for instance, when we consider the Bohmian
formulation of quantum mechanics \cite{A,sanz-AOP}, which allows us to
introduced suitable well-defined trajectories into the domain
of quantum optics without violating any fundamental principle.
Bearing this in mind, here we address the question of whether the set
of trajectories determined by the Bohmian picture can still provide a
reliable representation of polarization for quantum light states as an
ensemble of electric-field trajectories.
This would provide us with a rather intuitive model to understand
quantum light closer to the original idea of polarization.

Recently the trajectories described by the electric field of
one-photon two-mode states have been determined by following this approach
\cite{LS13}.
Here we extend the analysis to more relevant examples of classical-like
two-mode field states, namely as Glauber and SU(2) coherent states
\cite{CS,SU2-1,SU2-2}.
We specifically focus on this kind of states because a priori one
might naively expect that they would constitute the appropriate arena
to disclose the statistical description of polarization we are looking
for.
Surprisingly, we have found that for all these examples of
classical-like light the electric-field trajectories are
clearly incompatible with classical electrodynamics.
Actually, for the SU(2) states we have found that they are far away
from even resembling ellipses.
For completeness, these results are compared to polarization
trajectories associated with highly nonclassical stationary field,
such as $N00N$ states.

The two-dimensional harmonic oscillator has been considered as
a working model in previous Bohmian analyses \cite{holland-bk,A2},
although it is typically associated with matter waves.
Notice that the Bohmian approach is traditionally linked to quantum
mechanics, and only recently it has also been used in problems involving
electromagnetic fields (photons) -- even if in the 1970s and 1980s a few
authors already considered the possibility to extend the Bohmian approach
to electromagnetic fields.
However, in the area of quantum polarization, which we consider here, it
has been little exploited as an analysis working tool.
Here we report on an intriguing result, namely that Bohmian trajectories may
reveal nonclassical polarization dynamics displayed by polarization field
states that universally regarded as classical.

This work has been organized as follows.
In Section~\ref{sec2} we present the prescriptions to define
polarization trajectories as well as a discussion on the influence of
singular points on the polarization dynamics.
The dynamics associated with the polarization trajectories for
coherent, classical-like states are reported and discussed in
Section~\ref{sec3}, while in Section~\ref{sec4} we deal with the
counterpart for nonclassical $N00N$ states.
To conclude, a series of final remarks are summarized in
Section~\ref{sec5}.


\section{Polarization trajectories}
\label{sec2}


\subsection{Wave function for the electric field}
\label{sec21}

Usually the Bohmian formulation of quantum mechanics is applied to the
evolution of a particle in its position (configuration) space, which
involves the wave function in the corresponding coordinate
representation. In this work we make an effective transfer of this
formulation to the evolution of a two-mode electric field, which
involves the corresponding wave
function for the field variables.
Fortunately, such a transition is quite simple and straightforward,
since each field mode is formally equivalent to a mechanical harmonic
oscillator.

To take advantage of that equivalence in the simplest terms, we recall that a one-dimensional
mechanical harmonic oscillator of mass $M$ with Hamiltonian
\begin{equation}
H = \frac{p^2}{2M}+\frac{1}{2} k q^2,
\end{equation}
with $q$ and $p$ being its position and linear momentum.
This Hamiltonian can be quite conveniently described in terms of the
dimensionless creation $b^\dagger$ and destruction $b$ operators,
defined as
\begin{equation}
\label{b}
b = \sqrt{\frac{k}{2 \hbar \omega}}\ \! q + i \frac{p}{\sqrt{2 \hbar \omega M}},
\end{equation}
with $ \omega = \sqrt{k/M}$, and that satisfy the commutation relation
$[b, b^\dagger] =1$, so that $H = \hbar \omega (b^\dagger b + 1/2)$.

Likewise, a quantum one-mode electric field of frequency $\omega$ can be readily described by
the complex amplitude operator $a$ as $E \propto a \exp(-i \omega t )$ in the complex representation.
The operators $a$ and $a^\dagger$ satisfy the same commutation relation of
$b$ and $b^\dagger$, this
is $[a,a^\dagger] = 1$. Regarding the real and imaginary parts of $E$, the following quadrature
operators are defined
\begin{equation}
\label{a}
X = \frac{1}{\sqrt{2}} \left ( a + a^\dagger \right ) , \quad
Y = \frac{i}{\sqrt{2}} \left ( a^\dagger - a \right ) ,
\end{equation}
which satisfy the commutation relation  $[X,Y] = i$.
These operators can be regarded, respectively, as the field
counterparts of the mechanical position $q$ and linear momentum $p$
operators.
This allows us to introduce the quadrature representation of any one-mode field state
$|\psi \rangle$ as  $\psi (x ) = \langle x | \psi \rangle$ in terms of the eigenstates $| x \rangle$
of the quadrature operator $X$, where $X | x \rangle = x | x \rangle$. In this representation the
quadrature operators become $X \psi (x ) = x \psi (x ) $ and $Y \psi (x ) = - i \partial \psi (x) /
\partial x$.

After Eqs.~(\ref{b}) and (\ref{a}) the equivalence between the mechanical oscillator and the field
mode can be carried out in very simple terms if we take units in which $\hbar = m =\omega = 1$.
For example, the free-field Hamiltonian reads $ H = a^\dagger a + 1/2$. More importantly, we can
easily construct the wave functions for the number states $| n \rangle$ and the coherent states
$|\alpha \rangle$, defined by the eigenvalue equations
\begin{equation}
\label{eig}
a^\dagger a | n \rangle = n | n \rangle, \qquad a | \alpha \rangle = \alpha | \alpha \rangle,
\end{equation}
in terms of their mechanical counterparts after Eqs.~(4.1.32) and  Eqs.~(4.3.41) in Ref. \cite{GZ}
as
\begin{equation}
\label{psin}
\psi_n (x) =  \langle x | n \rangle
  = \frac{1}{\sqrt{2^n n! \sqrt{\pi}}}\ \! H_n (x) e^{-x^2/2},
\end{equation}
where $H_n$ are the corresponding Hermite polynomials, and
\begin{equation}
\label{psia}
\psi_\alpha (x) =  \langle x | \alpha \rangle = \left ( \frac{1}{\pi} \right )^{1/4} e^{-(x-\tilde{x} )^2 /2}
e^{i \tilde{y} x}  ,
\end{equation}
where $\sqrt{2} \alpha = \tilde{x} + i \tilde{y}$. Both expressions can be easily derived from
the the eigenvalue equations (\ref{eig}) and the differential form of the couple-amplitude operator
$a \rightarrow ( x + \partial/\partial x )/\sqrt{2}$ in the quadrature representation $\psi(x)$.

After  this transformation from quantum mechanics to quantum optics,
the only care to be taken is to remember that here $x$ does not
represent a position, but the electric field in the form of the field
quadrature $X$. This allows us the following fruitful translation to polarization of the Bohmian formulation
of the evolution of a two-dimensional harmonic oscillator.


\subsection{Polarization guidance equation}
\label{sec21}

In analogy to the Bohmian formulation of quantum mechanics or, in short,
Bohmian mechanics \cite{A}, polarization trajectories described by the
transverse electric field for two-mode harmonic light can be obtained by
solving the guidance equation \cite{LS13}
\begin{equation}
 \dot{\bm{x}} = {\bf \nabla} S ,
 \label{ge}
\end{equation}
where $\bm{x} = (x_1, x_2)$ denote the real, transverse components of
the electric-field strength in Cartesian coordinates, $S$ is the phase
of the field-state wave function in quadrature representation, and the
gradient ${\bf \nabla} S$ is taken with respect to $\bm{x} = (x_1, x_2)$.
Notice that, as in the usual Bohmian formulation, we have recast the
electric field wave function in polar form,
\begin{equation}
 \psi (\bm{x},t ) \equiv \langle \bm{x} | \psi (t) \rangle
  = \left | \psi (\bm{x},t ) \right | e^{i S (\bm{x},t )} ,
\end{equation}
where we have assumed the field state $|\psi(t)\rangle$ to be pure, and
$|\bm{x}\rangle = | x_1 \rangle | x_2 \rangle$ represents the eigenstates
of the corresponding quadrature operators.

Notice that because light polarization just describes the time-evolution
of the electric field, its configuration space is given by the electric-field
variables $(x_1,x_2)$, which play the same role as the coordinates
$\bm{r}$ in the case of particle dynamics. Once the general guidance
equation is established, sets of polarization trajectories are determined
by plugging the corresponding quantum field state (its phase) into this
equation, and then solving it for some particular set of initial conditions,
as in classical mechanics.

It is worth pointing out that in the quantum case all the information
about the polarization state is encoded in the scalar wave function
$\psi(\bm{x},t)$.
More importantly, since $\psi(\bm{x},t)$ represents a probability
amplitude, its phase has no classical analog.
Therefore everything about the polarization trajectories relies on a
nonclassical object, and hence we should expect that most conclusions
derived from the phase of $\psi (\bm{x},t)$ will have no classical
counterpart at all.


\subsection{Equilibrium points}
\label{sec22}

Except for the Glauber coherent states, here we have essentially focused
on stationary states, and hence the topology of their phase in the
polarization configuration space is going to be time-independent.
This means that any motion will be associated with this topology rather
than with the time-evolution of the phase gradient, as shown in \cite{LS13}.
In this case, the mathematical framework of the stability theory is of
much interest, for it may allow us to elucidate dynamical properties
by identifying possible equilibrium points \cite{jordan-bk}.
This points are going to determine the behavior of the polarization
trajectories in their vicinity and, therefore, the general dynamical
landscape associated with each quantum state.

The nodes or zeros of the wave function, where $\psi (\bm{x}) = 0$
and the phase $S$ is undefined, constitute the first kind of candidate
to equilibrium point.
As is well known \cite{berry1,berry2,berry3}, nodes or phase
singularities organize the global spatial structure of the flow of an
optical field.
This follows from Stoke's theorem, which states that unless the curl
of certain vector field is 0 within a certain region (irrotational
flow), the line integral around a closed loop enclosing such a region
(i.e., the circulation of such a vector field) will be nonzero.
If this vector field is identified with the phase gradient, then we
know that this quantity will be invariant under the addition to the
phase of any integer multiple of $2\pi$.
Consequently, if the curl of $\nabla S$ is nonzero, its circulation
will be quantized.
This is precisely what we observe in the case of Bohmian trajectories
that whirl around a node of the wave function
\cite{hirsch-1,hirsch-2,asanz-v1,asanz-v2},
where the quantization is in terms of integer multiples of $2\pi \hbar$.
Of course, this also holds for the polarization trajectories that we
are dealing with here, since
\begin{equation}
 \oint \dot{\bm{x}} \cdot  d\bm{x}  = \oint dS = 2 \pi \sigma .
\end{equation}
As it can be inferred from the latter integral, the presence of zeros in
the wave function allows the introduction of a circulation number or
topological charge, $\sigma$.
This number has to be an integer, since the line integral provides the
change experienced by the phase after an excursion returning to the
original point \cite{Be,merzbacher,riess-1,riess-2}.
Or, in topological terms, it accounts for the number of jumps between
different equivalent points of the Riemann surface described by the
logarithm of the wave function. In all cases examined in this work, the
trajectories around the zeros will be nearly circular in a neighborhood
of the node, giving rise to a vortical dynamics
\cite{hirsch-1,hirsch-2,asanz-v1,asanz-v2,Be}.
It is worth noting that in quantum  mechanics, it was Dirac who first
noticed this effect \cite{dirac}, suggesting the existence of magnetic
monopoles.
The concept of magnetic monopole has been further developed
in the literature within the grounds of quantum hydrodynamics
\cite{bialynicki}.
On the other hand, recently it has also been possible to recreate in
laboratory conditions Dirac's monopoles making use of the properties
displayed by different materials, such as crystals made of spin ice
\cite{monopole-2009-1,monopole-2009-2,monopole-2009-3,monopole-2009-4}
or Bose-Einstein condensates of rubidium atoms \cite{monopole-2014}. Or course,
strictly speaking, these are not elementary monopoles, but
quasi-particles arising as an emergent phenomenon associated with a
collective behavior, which display analogous properties to the
hypothesized Dirac monopole.

Critical or stationary points, i.e., points at which all partial
derivatives of a given function are zero, constitute the second kind of
equilibrium point that we may identify.
In our particular context, stationary points $\bm{x}_s$ will produce
a vanishing phase gradient, i.e., ${\bf \nabla} S = 0$.
That is, given the guidance equation (\ref{ge}), we will find
\begin{equation}
 \frac{d^r \bm{x}}{dt^r} \bigg\arrowvert_{\bm{x} = \bm{x}_s} = {\bf 0}
\end{equation}
for all $r > 0$. In all the cases examined in this work, the trajectories near these
points are hyperbolic, with the corresponding value of $\bm{x}_s$
being a saddle point of the velocity field ${\bf \nabla} S$.
We have found no maxima or minima of $S$, which would lead respectively
to sinks and sources of trajectories.

It is worth noticing that both nodes and stationary points are zeros of
the current density, $\bm{j} = |\psi |^2 \nabla S = {\rm Im} \left
( \psi^\ast {\bf \nabla} \psi \right )$.
Nonetheless, the role played by these two types of equilibrium points
is different.
The asymptotically stable and unstable branches associated with the
stationary points define separatrices around the nodes, which determine
domains with different dynamical behavior.
In particular, the direction of the flow around the nodes changes the
sign when one passes from the domain of one of these nodes to another
adjacent one.
As it will be seen below, in some cases these domains are included
within a larger domain with a preferential flow direction, while in
others the full configuration space is totally divided in domains
without enabling the appearance of larger domains.


\section{Field trajectories for classical-like states}
\label{sec3}


\subsection{Glauber coherent states}
\label{sec31}

Two-mode quadrature coherent states, typically known as Glauber coherent
states and denoted as $|\alpha_1,\alpha_2 \rangle$, constitute the
paradigm of classical light. Their electric-field wave function after Eq.~(\ref{psia})
is in this two-mode field scenario:
\begin{equation}
 \psi (\bm{x},t) \propto e^{-\left ( \bm{x} -  \bm{\tilde{x}} \right)^2 /2}
  e^{ i \bm{\tilde{y}} \cdot \bm{x} } ,
 \label{Gcwf}
\end{equation}
where $\bm{\tilde{x}}$ and $ \bm{\tilde{y}}$ are real two-dimensional vectors
defined according to the relation $\sqrt{2} \bm{\alpha} e^{-it} = \bm{\tilde{x}}+ i \bm{\tilde{y}}$,
being $\bm{\alpha} = ( \alpha_1, \alpha_2 )$. Thus these vectors evolve in time as
\begin{equation}
\tilde{x}_\ell = \sqrt{2} | \alpha_\ell | \cos (t - \delta_\ell) , \quad
\tilde{y}_\ell = - \sqrt{2} | \alpha_\ell | \sin (t - \delta_\ell) ,
\end{equation}
with $\delta_\ell = \arg{\alpha_\ell}$ for $\ell=1,2$. Since the phase is $S = \bm{\tilde{y}} \cdot \bm{x}$
the guidance equation is simply $\dot{\bm{x}} = \bm{\nabla} S =  \bm{\tilde{y}}$, which can be
easily solved analytically to give
\begin{equation}
 x_\ell (t) = x_\ell (0) + \sqrt{2} | \alpha_\ell | \cos (t - \delta_\ell) - \sqrt{2}| \alpha_\ell | \cos \delta_\ell ,
 \label{tr1}
\end{equation}
this is to say
\begin{equation}
 \bm{x}  (t) -  \bm{x} (0) = \bm{\tilde{x}} (t) - \bm{\tilde{x}} (0) .
 \label{tr2}
\end{equation}

\begin{figure}[t]
 \begin{center}
 \includegraphics[width=8cm]{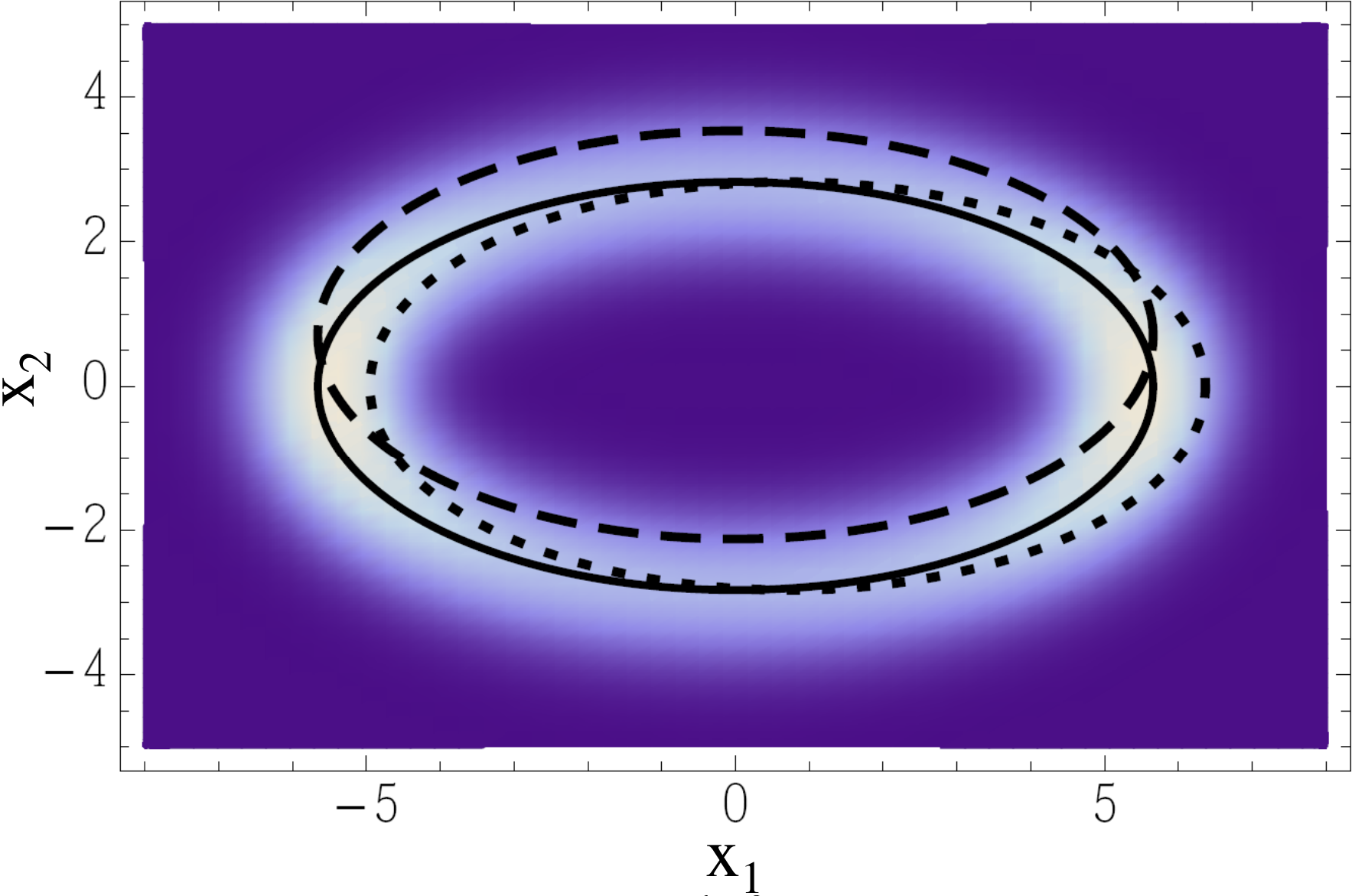}
 \caption{\label{fig1}
  Polarization trajectories for a two-mode coherent state with
  $\alpha_1 = 4$ and $\alpha_2 = 2i$.
  The contour-plot represents the probability density associated with
  the one-cycle averaged probability distribution for the electric field
  $\overline{P} (\bm{x})$.}
 \end{center}
\end{figure}

In Fig.~\ref{fig1} we have plotted three polarization trajectories for
a Glauber coherent state by considering different initial conditions in
Eq.~(\ref{tr1}).
The solid line represents the most probable trajectory, starting at
$t=0$ at the maximum of (\ref{Gcwf}), i.e., $\bm{x}(0) =
\bm{\tilde{x}}(0)$ so that $\bm{x}(t) = \bm{\tilde{x}}(t)$; the
other two trajectories (dashed and dotted lines) start from points at
one standard deviation from the maximum.
The background represents a contour-plot of the probability density
associated with the one-cycle averaged probability distribution for the
electric field
\begin{equation}
 \overline{P} (\bm{x} ) \propto
  \int_0^{2 \pi} \left | \psi \left (\bm{x} ,t \right ) \right |^2 dt .
\end{equation}

As happens with the Bohmian trajectories of a quantum harmonic
oscillator, here we also notice that any trajectory associated with a
Glauber state displays the same topology and keeps a constant distance
with respect to the most probable one, as it is inferred from
Eq.~(\ref{tr2}).
In this case, this topology coincides with the polarization ellipse of
a classical harmonic wave with complex-amplitude vector $\bm{\alpha}$.
Nonetheless, only the most probable trajectory (the solid line in
Fig.~\ref{fig1}) is centered at the origin; any other trajectory will
be slightly displaced, as mentioned before (see dashed and dotted lines
in Fig.~\ref{fig1}).
Despite coherent states are regarded as typical examples of
classical-like light, such a displacement constitutes an important
difference with respect to what one would expect from classical
electrodynamics, namely zero displacement (i.e., concentric
trajectories).


\subsection{SU(2) coherent states}
\label{sec32}

The two-mode Glauber coherent states define another interesting family of classical-like
states regarding polarization, namely the SU(2) coherent states. These states arise after
recasting the two-mode Glauber coherent states as \cite{SU2-1,SU2-2}
\begin{equation}
 \label{GvSU}
 |\alpha_1 , \alpha_2 \rangle = e^{-|\alpha|^2/2}
 \sum_{n=0}^\infty \frac{| \alpha |^n e^{in \delta} }{\sqrt{n!}} |n,\Omega\rangle .
\end{equation}
Here $|n,\Omega\rangle$ denotes the SU(2) coherent state with $n$ photons,
which reads explicitly as
\begin{equation}
\label{SUn}
 |n,\Omega\rangle = \sum_{m=0}^n \begin{pmatrix}  n \cr m \end{pmatrix}^{1/2}
  \cos^m \frac{\theta}{2} \sin^{n-m}
  \frac{\theta}{2} e^{-im \delta } | m, n-m \rangle ,
 \end{equation}
where $|m,n-m\rangle$ are two-mode photon-number states, and (assuming $\alpha_1$ real without loss of generality)
\begin{equation}
 \alpha_1 = | \alpha | \cos \frac{\theta}{2} , \qquad
 \alpha_2 = | \alpha | \sin \frac{\theta}{2}\  e^{-i\delta} .
\end{equation}
It is worth noting that the polarization state, as given by the Stokes
parameters, is the same for the Glauber coherent states and all the
SU(2) coherent states in Eq.~(\ref{GvSU}).

The wave function $\psi (\bm{x},t)$ accounting for SU(2) coherence
states with $n$ photons, given by Eq.~(\ref{SUn}), reads after Eq.~(\ref{psin}) as
\begin{equation}
 \label{wfcn}
 \psi (\bm{x},t) \propto
  \sum_{m=0}^n \frac{\alpha_1^m \alpha_2^{n-m}}{m! (n-m)!}
   H_m (x_1) H_{n-m} (x_2) e^{-\bm{x}^2 /2} e^{-int} .
\end{equation}

Here, the only analytical solution to the guidance equation holds in
the particular case of $|\alpha_1| = |\alpha_2|$ and $\delta =
\pm \pi/2$, for which we have $\psi (\bm{x}) \propto
\left( x_1 \pm  i x_2 \right)^n e^{-\bm{x}^2 /2}$.
In this case, all trajectories are circles and there is only one node
at the origin with charge $\sigma = \pm n$, the sign depending on the helicity of
the classical polarization ellipse.

\begin{figure}[t]
 \begin{center}
 \includegraphics[width=8.25cm]{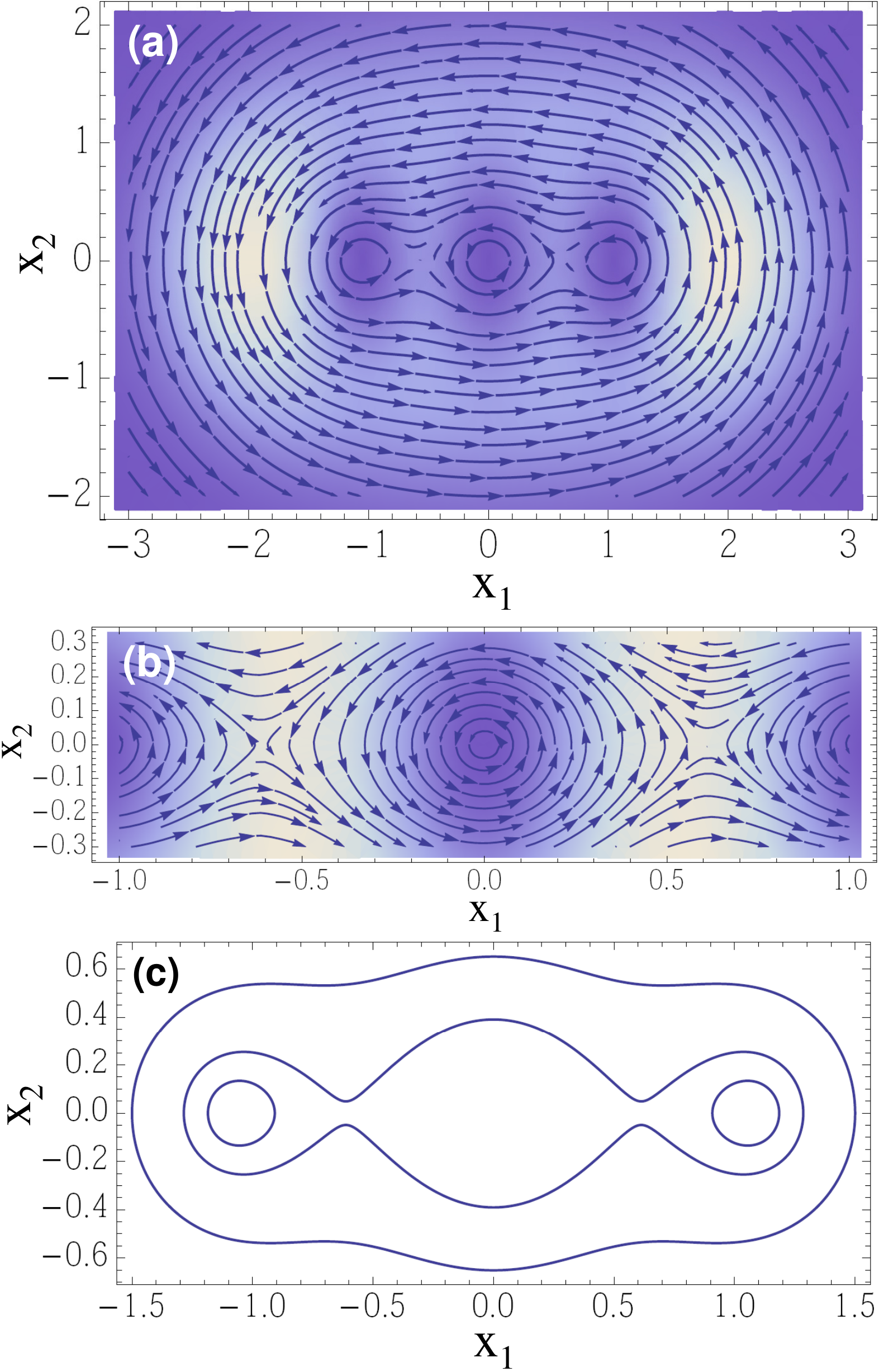}
 \caption{\label{fig2}
  (a) Streamlines illustrating the trajectory dynamics associated with
  an SU(2) coherent state with $\alpha_1=4$, $\alpha_2=2i$, and $n=3$.
  The contour-plot represents the probability density $|\psi(\bm{x},t)|^2$
  of the electric field.
  (b) Enlargement of panel (a) to show the dynamics around the
  central node and the two adjacent hyperbolic stationary points.
  (c) Four polarization trajectories showing the incompatibility with
  the classical electrodynamics of a freely evolving two-mode harmonic
  electric field.}
 \end{center}
\end{figure}

This coincides exactly with the circular polarization associated with
the complex-amplitude vector $\bm{\alpha}$.
For any other general case, the trajectories shown in Fig.~\ref{fig2}(a)
provide an idea of the general trend.
These trajectories, displayed in the form of streamlines (arrows
indicate the directionality of the motion), correspond to an SU(2)
coherent state with $\alpha_1 = 4$, $\alpha_2 = 2i$, and $n=3$;
the contour-plot represents the probability density
$|\psi (\bm{x},t)|^2$ associated with the coherent state considered.
This example illustrates without loss of generality the results we have
found for all cases examined, specifically that there are $n$ nodes
located along the major axis of the classical ellipse associated with
the complex vector $\bm{\alpha}$.
In the vicinity of the nodes, the trajectories are nearly circular
\cite{Be}, as can be better seen in the enlargement around the central
node provided in Fig.~\ref{fig2}(b).
The three nodes have the same topological charge $\sigma = + 1$.
Between any two consecutive nodes, along the line connecting them,
there are $n-1$ hyperbolic stationary points \cite{jordan-bk}.
In the vicinity of these points, the trajectories display a hyperbolic
topology with identical semi-axes [see Fig.~\ref{fig2}(b)].

The trajectory that passes just through the two stationary points is a
separatrix, which separates the three dynamical domains associated with
each note from a single outer domain, where the trajectories move around
the all three nodes.
Actually, far from the nodes the trajectories approach circles.
This can be readily shown analytically by considering the approximation
$H_n(x) \approx (2x)^n$ for large $x$, and substituting it into the wave
function (\ref{wfcn}), which yields
\begin{equation}
 \psi(\bm{x}) \propto \left (\bm{\alpha} \cdot \bm{x} \right)^n
   e^{-\bm{x}^2 /2}  ,
\end{equation}
and hence $\dot{\bm{x}} \propto (x_2, - x_1 )$, $\dot{\bm{x}} \cdot \bm{x} =0$,
and $| \bm{x} | = {\rm constant}$ along each trajectory, which define a
circular motion.
Of course, the rotation of the outer trajectories the three central
domains can be associated with the motion around a single effective
node of charge $|\sigma|=n$.
As an illustration of the extremely streaking behavior displayed by
the polarization trajectories associated with these classical-like
polarization states, a representative set of them is shown in
Fig.~\ref{fig2}(c).
As is apparent, the behavior exhibited by all these trajectories is
clearly incompatible with the classical electrodynamics corresponding
to a freely evolving two-mode harmonic electric field.


\section{Nonclassical field: $N00N$ states}
\label{sec4}

For the sake of comparison, we will also briefly consider a
paradigm of nonclassical state, namely a $N00N$ state.
These states constitute the polarization analog of Schr\"odinger cat
states or coherent superpositions of distinguishable states.
In the photon-number basis they read
\cite{N00N-1,N00N-2,N00N-3,N00N-4} as
\begin{equation}
 |\psi \rangle \propto \alpha_1 | n, 0 \rangle + \alpha_2 | 0,n \rangle .
\end{equation}
This can be regarded as an alternative quantum version of the coherent
superposition of two orthogonal oscillations, which is the actual origin
of polarization.
The corresponding wave function is
\begin{equation}
 \psi(\bm{x}) \propto
  \left [ \alpha_1 H_n ( x_1 ) + \alpha_2 H_n (x_2) \right ]
  e^{-\bm{x}^2/2} .
\end{equation}

\begin{figure}[t]
\begin{center}
\includegraphics[width=7cm]{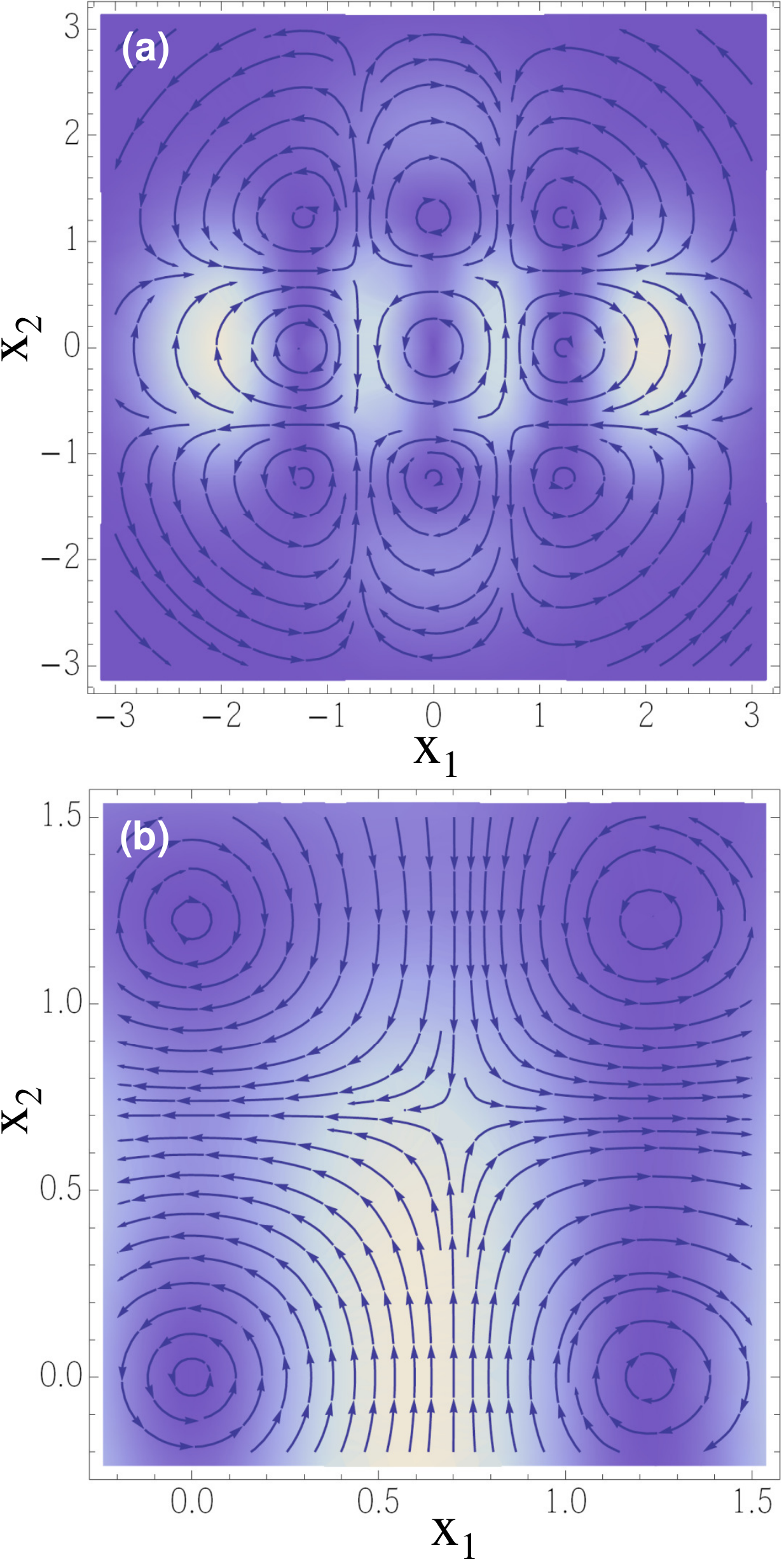}
 \caption{\label{fig3}
  (a) Streamlines illustrating the trajectory dynamics associated
  with a $N00N$ state with $\alpha_1=4$, $\alpha_2=2i$, and $n=3$.
  The contour-plot represents the probability density
  $|\psi(\bm{x},t)|^2$ of the electric field.
  (b) Enlargement of panel (a) to show the dynamics around
  one of the hyperbolic stationary points and the corresponding four
  adjacent nodes.}
 \end{center}
\end{figure}

A global picture of the dynamics for $N00N$ states is provided in
Fig.~\ref{fig3} (left) in terms of streamlines.
To compare with the case analyzed in Section~\ref{sec3}, the values
$\alpha_1 = 4$, $\alpha_2 = 2i$, and $n=3$ have been considered again.
An enlargement showing the dynamical details in the vicinity of one of
the hyperbolic stationary points and the adjacent nodes can be seen in
the right panel.
The nodes form an array of $n\times n$ domains, where the trajectories
are nearly circles with $\sigma=\pm 1$.
In this case, the sign of $\sigma$ is always opposite for nearest neighbors.
On the other hand, the stationary points form a $(n-1)\times(n-1)$ array,
such that the corresponding separatrices divide the configuration space
into isolated domains regardless of how far we find from the nodes.
These points are located along the diagonals connecting nodes with the
same sign of $\sigma$ and, again, the trajectories in their vicinity display
a hyperbolic topology.


\section{Final remarks}
\label{sec5}

We have addressed a Bohmian approach to light polarization in quantum
optics \cite{A,sanz-AOP,LS13} by computing the trajectories described
by the electric field of some classical-like two-mode states. In general, we
have noticed that for Glauber coherent states all trajectories have the same
elliptic form of the mean field, even if they are not centered at the origin.
To some extent, this was an expected result.
However, for SU(2) coherent states, the corresponding trajectories are
further away from being ellipses.
This can be ascribed to the fact that, although SU(2) coherent states are
stationary, they are still able to exhibit a trajectory dynamics due to
local phase variations, i.e., to a purely geometric origin, as previously
shown in the case of single-photon superpositions \cite{LS13}.

Now, what is really quite remarkable here is the fact that
polarization trajectories may display a dynamics far beyond the expected
classical elliptical trajectories, even in the case of polarization field
states that are universally regarded as classical.
Instead, the results reported show that many different trajectories, with
very different topologies, are compatible with every single state, which
is in compliance with some recent quantum-polarization approaches, where
the degree of polarization can never reach unity
\cite{S3,rap1,rap3,rap4}.

In principle, these new results should be observable in practice in
virtue of the close relationship between the Bohmian picture of quantum
dynamics and the concept of weak measurement \cite{wm1,wm2,wm3}.
Notice that this connection has already been proven experimentally in
benchmark experiments \cite{be1,be2}.
Furthermore, it has also been adapted to the case where the Bohmian
trajectories hold in the field-quadrature space by means of the homodyne
scheme, as shown in Ref.~\cite{FF12}.

As shown here, the strange dynamics reported for SU(2) coherent states
is primarily determined by the equilibrium points of the electric-field wave
function, in particular a web of vortices. Analogously, this geometrical nature
has also been observed in nonclassical light. This naturally leads to the issue
of Bohmian chaos \cite{ABM}, which is absent in all cases analyzed here,
because vortices need to be evolving in time.
In 1995 Parmenter and Valentine \cite{PV} showed that just a linear
superposition of eigenstates of a 2D anisotropic harmonic oscillator
might lead to chaos under some specific conditions, an idea that Makowski
and Frackowiak \cite{MF} further analyzed later on in 2001, identifying
the ``simplest non-trivial model of chaotic Bohmian dynamics''.
Nonetheless, the link between Bohmian chaos and vorticality was first
established by Frisk in 1997 \cite{frisk}, and more recently Wiskniacki,
Pujals and Borondo \cite{WP,WPB} have found out that, in particular, it
is the movement of vortices what induces the appearance of chaos, which
explains why there are no signatures of chaos in our case.
From a dynamical viewpoint, the states analyzed are all stable, although
a slight perturbation would lead to motion of the observed saddle points
and, therefore, to the appearance of chaos, although this is a subject
that goes beyond the scope of the current work.

The above results, in contradiction with the type of dynamics that
one would expect in principle from classical electrodynamics, constitute
a quite remarkable issue, since such field states are universally regarded
as classical-like concerning polarization.
Nonetheless, there are some quantum approaches where
these states also display nonclassical polarization features, as discussed in
\cite{cnc-1,cnc-2,cnc-3}. In this regard, a natural question that arises here is whether
there is any relationship between the manifestation of nonclassical polarization in
these approaches and the one here discussed within the Bohmian framework. The
answer is positive, there being a straightforward link between them. In terms of a
mechanical-like language, the phase gradient in Eq.~(\ref{ge}) provides the local
value of a linear momentum. This can be suitably expressed as a local mean value
of the momentum either via Wigner-Moyal phase-space distributions or
Terletsky-Margenau-Hill ones \cite{lmv-1,lmv-2,lmv-3,lmv-4,lmv-5,WvT}, which
are the ones displaying nonclassical behavior in \cite{cnc-1,cnc-2,cnc-3}.
Trajectories displaying strange behaviors might be then regarded as the
result of quantum polarization distributions incompatible with classical
physics.

From the above comments, it is clear that Glauber and SU(2) coherent
states must be separately analyzed.
The Wigner distribution for Glauber coherent states is classical
(it is everywhere positive definite) and, consequently, one should
go to nonlinear functions of the trajectories. This is because nonlinear
local moments are related exclusively to
Terletsky-Margenau-Hill \cite{WvT}, which is nonclassical for
Glauber coherent states \cite{cnc-1,cnc-2,cnc-3}.
Regarding SU(2) coherent states, their characteristic trait is the
presence of vortices governing the topology of the trajectories.
These vortices arise when the amplitude of the wave function vanishes.
This vanishing implies that both the Wigner and the
Terletsky-Margenau-Hill distributions will be negative
definite in regions around vortices.
Roughly speaking, this means that the trajectories orbiting the
vortices should be influenced by the nonclassical negative values of
these distributions.

To conclude,  we would like to stress the fact that the definition
of the electric-field trajectories has no straightforward classical
counterpart. There seems to be no simple classical analog for
the phase of the electric-field wave function.


\section*{Acknowledgements}

We acknowledge financial support from the Ministerio de Econom{\'\i}a y Competitividad (Spain)
under Projects Nos.~FIS2012-35583 (A.L.) and FIS2011-29596-C02-01 (A.S.), and a ``Ram\'on y Cajal''
Research Fellowship [Ref.~RYC-2010-05768 (A.S.)], as well as from the 
Comunidad Aut\'onoma de Madrid research consortium QUITEMAD+ grant S2013/ICE-2801 (A.L.).


\end{document}